# Enzyme Kinetics: A critique of the quasi-steady-state approximation

Kamal Bhattacharyya[1] and Sharmistha Dhatt

*Department of Chemistry, University of Calcutta, Kolkata 700 009, India*

**Abstract**

The standard two-step model of homogeneous-catalyzed reactions had been theoretically analyzed at various levels of approximations from time to time. The primary aim was to check the validity of the quasi-steady-state approximation, and hence emergence of the Michaelis-Menten kinetics, with various substrate-enzyme ratios. But, conclusions vary. We solve here the desired set of coupled nonlinear differential equations by invoking a new set of dimensionless variables. Approximate solutions are obtained via the power-series method aided by Padè approximants. The scheme works very successfully in furnishing the initial dynamics at least up to the region where existence of any steady state can be checked. A few conditions for its validity are put forward and tested against the findings. Temporal profiles of the substrate and the product are analyzed in addition to that of the complex to gain further insights into legitimacy of the above approximation. Some recent observations like the 'reactant stationary approximation' and the notions of different timescales are revisited. Signatures of the quasi-steady-state approximation are also nicely detected by following the various reduced concentration profiles in triangular plots. Conditions for the emergence of Michaelis-Menten kinetics are scrutinized and it is stressed how one can get the reaction constants even in the absence of any steady state.

[1]pchemkb@yahoo.com; pchemkb@gmail.com (Corresponding author)

# 1. INTRODUCTION

The standard two-step model of homogeneous-catalyzed reactions leads to a set of coupled differential equations. Several interesting features[1–5] of nonlinearity in such reactions involve biochemical systems, either in isolation or as part of complex reaction networks. Therefore, simplifying assumptions are often made for the solutions. The most popular and useful result of these endeavors is the Michaelis–Menten (MM) form[6–8], particularly relevant to enzyme kinetics. One assumes here that the concentration of enzyme-substrate complex remains approximately constant over a considerable time span after a short transient. This is commonly known as the quasi-steady-state approximation (QSSA). It has been customary to test QSSA by choosing large substrate-enzyme ratios in keeping with *in vivo* studies. Therefore, one is led to believe, along with many authors[9–17] that the standard QSSA (s–QSSA) is valid only when the enzyme concentration is small enough, though the range of validity of the MM region is widened[13]. On the other hand, a number of studies[2, 18–25] considered moderate-to-large enzyme-substrate ratios and found QSSA regions there too, under specific circumstances. These are relevant to interesting *in vitro* studies. Such endeavors, without any restriction on substrate-enzyme ratios, look for the applicability of total QSSA (t–QSSA). Experimental relevance of the MM kinetics in these situations is also available[4].

The two-step model corresponds to the reaction scheme

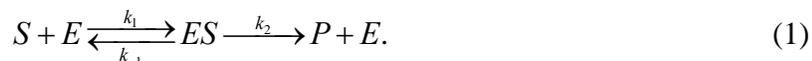

$$S + E \underset{k_{-1}}{\overset{k_1}{\rightleftarrows}} ES \xrightarrow{k_2} P + E. \tag{1}$$

Several features of QSSA have been noted on its basis. For example, Laidler[26] put forward certain conditions for the applicability of QSSA. Borghans *et al*[19] distinguished s–QSSA from t–QSSA, and also remarked on reverse QSSA (r–QSSA) when the enzyme–substrate ratio is large. The idea was extended by Tzafriri[22]; subsequent extensions[23, 24] followed. Various perturbation methods[10, 13, 18] have appeared with different scaled variables to understand QSSA. A nice summary of such works with further developments are available[27, 28]. A variation-iteration method due to He[29–31] has recently[32] been found effective over a range of certain parameter values. Legitimacy of the MM approximation via a stochastic algorithm has also been forwarded[15].

From an experimental point of view, however, one can follow not only the rate of product formation, but the temporal profiles of substrate and product also. Theoretical studies, on the other hand, are centered chiefly on the profile of the complex. Conditions for the validity of QSSA also vary. So, while it is tempting to explore better and precise conditions of the applicability of QSSA (or t–QSSA) from a different theoretical approach to the problem, one may also legitimately inquire whether its signatures exist in substrate–time and product–time plots. We also indicate the usefulness of triangular plots in deciphering the applicability of QSSA. A few other relevant queries include (i) relations among the maximum complex concentration, transient time and QSSA, if any, (ii) dependence of the steady–state region on the starting enzyme-substrate ratio[17], (iii) adequacy of the 'reactant stationary approximation' (RSA)[21, 33], (iv) relevance of two time–scales and their relations in s–QSSA[18 – 20, 22 – 24] and r–QSSA[19, 21], etc.

Another important area concerns the validity of the MM kinetics. It is generally believed that the same rests on the assumption of the QSSA; thus, when QSSA fails to be obeyed, the MM kinetics loses its footing. However, we like to scrutinize the role of MM kinetics separately, irrespective of whether QSSA becomes valid or not. That this is possible will become clear in due course. One can also get the reaction constants from appropriate plots. Indeed, this turns out to be the most important part of the present work. Such a study has a good bearing on an early work[34] that claimed unacceptability of MM kinetics for about 800 enzymes!

To inspect the questions posed above, we choose a considerably different route. Casting the relevant equations in terms of a new set of dimensionless variables, we employ the standard power-series method of solution, supplemented by the construction of suitable Padè approximants (PA). Indeed, this is one of the most straightforward schemes to handle the problem. The success, however, depends on the choice of variables. In this respect, our scaling scheme has been found to perform nicely[35]. It extends the region of validity of the initial dynamics to intermediate times in a very successful manner. Our endeavor does not depend on the magnitude of the enzyme-substrate ratio[36] either. Hence, all the three types of QSSA can be dealt with on equal footing.

## 2. THE METHOD

**2.1. Scaling.** Denoting the concentrations of complex ES, enzyme E, substrate S and product P respectively by $c$, $e$, $s$ and $p$, and their initial values by a subscript zero, we obtain from (1) the following differential equations:

$$ds/dt = -k_1 es + k_{-1} c, \qquad (2)$$

$$dc/dt = k_1 es - (k_{-1} + k_2)c, \tag{3}$$

$$de/dt = -k_1 es + (k_{-1} + k_2)c, \tag{4}$$

$$dp/dt = k_2 c. \tag{5}$$

In addition, we have two mass conservation equations

$$e_0 = e + c, \tag{6a}$$

$$s_0 = s + c + p. \tag{6b}$$

To solve the above equations, various sets of scaled variables have been employed (see, e.g., Fraser[27], Murray[37], and works quoted therein). Here, we employ the following dimensionless variables:

$$\alpha = e/e_0, \beta = s/e_0, \gamma = c/e_0, \delta = p/e_0, \tau = k_2 t. \tag{7}$$

The conservation equations (6) now read as

$$\alpha + \gamma = 1 \tag{8a}$$

$$\beta + \gamma + \delta = \beta_0 \tag{8b}$$

The primary kinetic equations, out of (2) – (5), then follow as

$$-d\gamma/d\tau = d\beta/d\tau + \gamma, \tag{9}$$

$$d\beta/d\tau = -K_1 \beta + (K_1 \beta + K_2)\gamma, \tag{10}$$

with the initial conditions

$$\alpha_0 = 1, \beta_0 = s_0/e_0, \gamma_0 = 0 = \delta_0. \tag{11}$$

The constants $K_1$ and $K_2$ in (10) are given by

$$K_1 = k_1 e_0 / k_2, K_2 = k_{-1}/k_2. \tag{12}$$

Thus, we could reduce the actual problem by choosing three variables and two constants. The usual strategy[37] has been to employ three variables and three constants.

**2.2. Series Expansions.** Note that a large $K_2$ implies the equilibrium approximation in MM kinetics that had been extended [7] to QSSA in the same context long back. The above system of non-linear equations (9) – (10), with the aid of (11), can be solved analytically using the standard power series method. Hence, we express the concentrations of the participating species in power series of $\tau$, *viz.*

$$\beta_\tau = \sum_{j=0} \beta_j \tau^j, \gamma_\tau = \sum_{j=0} \gamma_j \tau^j, \tag{13}$$

etc., insert them suitably into (9) and (10), and collect similar powers of $\tau$. Thus, the unknown parameters of the expansions in (13) are obtained in terms of $\beta_0$, $K_1$ and $K_2$. The other concentration terms can then be obtained simply by invoking (8). A few results of future interest are the following:

$$\beta_1 = -K_1\beta_0,$$
$$\beta_2 = K_1\beta_0(K_1\beta_0 + \beta_0 + K_2)/2. \tag{14}$$

$$\gamma_1 = K_1\beta_0,$$
$$\gamma_2 = -K_1\beta_0(K_1\beta_0 + \beta_0 + K_2 + 1)/2. \tag{15}$$

By using equations (8), one can get similar expansion coefficients for $\alpha$ and $\delta$. In case of $\delta$, a better alternative is to directly integrate (5) that now takes the form

$$d\delta/d\tau = \gamma. \tag{16}$$

Let us also note that, while $\gamma$ rises from zero linearly during the initial phase of the reaction, it would finally tend to zero again. Hence, there exists at least one maximum in $\gamma$ - $\tau$ plot. Indeed, one finds from (9) and (10) that

$$d\gamma/d\tau = K_1\beta - (K_1\beta + K_2 + 1)\gamma. \tag{17}$$

It shows, the point where $d\gamma/d\tau$ becomes zero is unique and at this point $\tau_c$ the value of $\gamma$ would read as

$$\gamma_c = K_1\beta_c/(K_1\beta_c + K_2 + 1). \tag{18}$$

Therefore, there appears yet another possibility of expansions like (13). If one obtains $\tau_c$ and the corresponding concentrations $\gamma_c$ and $\beta_c$, then the new pair of expansions takes the form

$$\beta_\tau = \sum_{j=0} \beta_{cj}(\tau - \tau_c)^j, \quad \gamma_\tau = \sum_{j=0} \gamma_{cj}(\tau - \tau_c)^j. \tag{19}$$

Putting (19) in (9) and (10), we obtain the first few terms as

$$\beta_{c0} = \beta_c,$$
$$\beta_{c1} = -\gamma_c,$$
$$\beta_{c2} = K_1\gamma_c(1-\gamma_c)/2; \tag{20}$$

$$\gamma_{c0} = \gamma_c,$$
$$\gamma_{c1} = 0,$$
$$\gamma_{c2} = -K_1\gamma_c(1-\gamma_c)/2. \tag{21}$$

We shall see the usefulness of these terms later.

**2.3. Numerical Stability.** An obvious problem with the expansions like (13) is their inability to yield reliable results for large $\tau$. We circumvent here this problem by constructing the PA[38, 39]. The PA has been found to be quite faithful in perturbation theory involving divergent Taylor expansions and quite a few other contexts (see, e.g., Dhatt and Bhattacharyya[40] and references quoted therein). Here, we construct three types of PA, the diagonal [$N/N$] ones, and the two nearest off-diagonal [$(N+1)/N$] and [$N/(N+1)$] varieties. The agreement among values of such varieties points to the adequacy of the scheme. More specifically, we have taken the first 21 terms in (13) to obtain the sequences of these approximants. They suffice our purpose[35] as long as $K_1$ and $K_2$ are not large enough. Otherwise, one has to routinely increase the number of terms in order to get gradually improved results, or over a wider range of time $\tau$ at a fixed accuracy level.

Another way to check the numerical stability of our computed data is to compare the left and right sides of (18) from the PA sequences for $\gamma$ and $\beta$ at $\tau = \tau_c$. Indeed, this is the point at which rate of product formation attains its maximum value and, therefore, it possesses an experimental relevance too.

By following the above two checks, we noted that one can go well beyond the region of adequacy of QSSA. It may be pointed out that $\tau_c$ exists irrespective of whether QSSA is satisfied or not. Hence, the quality of steady state can be nicely assessed, if there is any, once the numerical scheme is known to be stable.

A different kind of possibility of extending the temporal regime is to first get $\gamma_c$ and $\beta_c$ via (13) and then employ (19). The rest of the scheme proceeds as before. After matching the coefficients, in the way we arrive at (20) and (21), one can construct the types of PA quoted above. However, in the present work, we did not require any use of (19) for numerical purposes.

## 3. ANALYSIS

**3.1. Behavior of the concentration profiles.** We consider first the case of $\gamma$. In the small-$\tau$ regime, it turns out that

$$\gamma = \gamma_1 \tau + \gamma_2 \tau^2 + ..., . \qquad (22)$$

where the coefficients are given by (15). It shows the initial linear rise, with a slope of $K_1\beta_0$. After the transient time $\tau_c$, however, it is expressible as

$$\gamma = \gamma_c + \gamma_{c2}(\tau - \tau_c)^2 + ..., \qquad (23)$$

in view of (19) and (21). Accepting that the quick linear rise is opposed by the quadratic term in (22) to yield a maximum, one can write

$$\gamma_c \approx \gamma_1 \tau_c + \gamma_2 \tau_c^2 \qquad (24)$$

and it can be solved for $\tau_c$, yielding

$$\tau_c \approx \gamma_c / \gamma_1 = \beta_c / (\beta_0 (K_1 \beta_c + K_2 + 1)). \qquad (25)$$

Result (25) should be approximately true for small $\tau_c$.

Initial fall-off of $\beta$ is linear too, with a slope of $K_1\beta_0$. Moreover, if $K_1$ is small, which we shall later see to turn out as a condition for QSSA, one can write

$$\beta \approx \beta_0 + \beta_1 \tau \approx \beta_0 \exp[-K_1 \tau] \qquad (26)$$

over a good range. Now, if it so happens that $K_1 \ll 1$, and the transient phase (0 to $\tau_c$) is small, then the RSA[21, 33] follows. Another characteristic parameter of some use[13, 18 - 24] is $\tau_s$, the time required for maximum change in $\beta$. For $\beta_0 \ll 1$, the initial decay is very slow. Hence, from (26), on the basis of initial decay, $\tau_s$ is the lifetime. Thus, we have a different timescale

$$\tau_s = 1 / K_1. \qquad (27)$$

This attaches a physical meaning to $K_1$. Note, however, that $\beta_2$ tends to oppose the fall-off. Around $\tau_c$, on the other hand, we find

$$\beta = \beta_c - \gamma_c (\tau - \tau_c) - \gamma_{c2} (\tau - \tau_c)^2 + ... \qquad (28)$$

that reveals again a linear fall-off unless $|\gamma_{c2}|$ is large. We shall see later how this result becomes useful.

Turning attention to $\delta$, we notice from (15), (16) and (22) that

$$d\delta / d\tau = \gamma \approx K_1 \beta_0 \tau + \mathbf{O}(\tau^2). \qquad (29)$$

It tells, the initial rise of the product is always parabolic in time[41]. However, unless $\tau_c$ is large, the parabolic nature may not show up significantly. Again, from (16), (21) and (23), one arrives at the temporal behavior of the product beyond $\tau_c$ as

$$\delta = \delta_c + \gamma_c (\tau - \tau_c) + \gamma_{c2} (\tau - \tau_c)^3 / 3 + ... , \qquad (30)$$

showing a linear rise for small enough $|\gamma_{c2}|$.

**3.2. Workability of the QSSA.** It is now appropriate to remark on the conditions so far put forward concerning the workability of QSSA. One of the earliest ones is given by Laidler[26]. Stated in terms of our parameters, his four conditions are

$$(a)\ \beta_0 \gg 1;$$
$$(b)\ \beta_0 \ll 1;$$
$$(c)\ K_1\beta_0/(1+K_2) \ll 1; \qquad (31)$$
$$(d)\ K_1/(1+K_2) \ll 1.$$

Either of these is a necessary condition. Additionally, it is agreed that, if (31a) holds, then $\tau_c$ would be small[11, 26]. As stated earlier, most authors favor (31a) only. Some authors[16, 17] still maintain that QSSA would fail under condition (31b). A few other works[18, 20] replace (31a) by

$$K_1/(K_1\beta_0 + K_2 + 1) \ll 1, \qquad (32)$$

highlighting it as the sole criterion for the validity of QSSA. It has also been remarked[19] that QSSA is tenable even when $\beta_0 \approx 1$, but then the Michaelis constant $k_m$ should obey

$$k_m = e_0(1+K_2)/K_1 \gg 1. \qquad (33)$$

An extension[22–24] of the earlier work[19] revealed that (32) would apply if $\beta_0 \gg 1$; in the converse case, one has to ensure whether

$$K_1\beta_0/(K_1 + K_2 + 1) \ll 1 \qquad (34)$$

is satisfied and this condition validates QSSA. Let us remark here that conditions (32) and (34) may better be viewed as extensions of (31d) and (31c), respectively. A thorough check[35] shows, however, that none of these conditions (31), (32) and (34) withstand a close scrutiny.

In terms of timescales, another idea[18-24] is to check the ratio $\tau_s/\tau_c$. For $\beta_0 \gg 1$, QSSA (or, s-QSSA) is said to be valid if

$$\tau_s/\tau_c \gg 1, \qquad (35)$$

with

$$\tau_s = \beta_0 + (K_2+1)/K_1,\ \tau_c = 1/(K_1\tau_s). \qquad (36)$$

At the other extreme (r-QSSA) of $\beta_0 \ll 1$, condition (35) is replaced by

$$\tau_s/\tau_c \ll 1, \qquad (37)$$

with

$$\tau_c = 1,\ \tau_s = 1/K_1. \qquad (38)$$

Workability of such conditions will be surveyed in the next section. We only remark here that when $\beta_0 \gg 1$, $\beta_0 \approx \beta_c$ is obeyed and hence (25) agrees with $\tau_c$ in (36), while the expression for $\tau_s$ in (27) matches with the same in (38) under r-QSSA condition.

In the present work, it is transparent from (21) and (23) that, if $\gamma$ needs to retain an approximate constancy (*i.e.*, $\gamma \approx \gamma_c$) over a considerable range of $\tau > \tau_c$, as is required for QSSA, then we should have

$$|\gamma_{c2}/\gamma_c| \ll 1. \qquad (39)$$

But, (21) shows that a sufficient condition for (39) to hold is

$$K_1 \ll 1. \qquad (40)$$

It now also explains the adequacy of the constant-$\beta$ approximation[33] discussed below (26). More appropriately, however, (39) yields

$$|\gamma_{c2}/\gamma_c| = K_1(1-\gamma_c)/2 \ll 1. \qquad (41)$$

Thus, (41) turns out to be a condition for QSSA.

Using (18), one finds from (41) that

$$\lambda_c = K_1(K_2+1)/\left(2(K_1\beta_c + K_2+1)\right) \ll 1. \qquad (42)$$

For convenience, we call the quantity at the left side of (42) by $\lambda_c$. However, its value depends on $\beta_c$ that may not be known *a priori*. So, we also define a quantity $\lambda_0$ by

$$\lambda_0 = K_1(K_2+1)/\left(2(K_1\beta_0 + K_2+1)\right). \qquad (43)$$

Then, while $\lambda_c > \lambda_0$, if one can ensure that $\lambda_0 \ll 1$, one may not be far from the applicability of QSSA. We shall later check how such a condition performs. Condition (43) looks in part like (31*c*) or (31*d*) and partly like (32). One can conclude from (43) that (i) a very small $K_1$ is sufficient for QSSA, as found before, but (ii) if $K_1$ is not small enough, we can still satisfy inequality (43) by requiring that $K_1\beta_c \gg (K_2 + 1)$. Further, for large $\beta_0$, we may replace the preceding inequality by $K_1\beta_0 \gg (K_2 + 1)$. Thus, it is neither true[9–17] that QSSA is always valid for $\beta_0 \gg 1$, nor false[17,42] that QSSA is always invalid for $\beta_0 \ll 1$.

**3.3. Conventional MM Kinetics.** The popular representation[7,8] of the MM kinetics reads for the rate *r* as

$$r = r_M s/(s+k_m),\ r_M = k_2 e_0, \qquad (44)$$

obtained from the condition

$$dc/dt = 0. \qquad (45)$$

In (44), $r_M$ is the maximum rate. Sometimes, (44) is written more simply as (s-QSSA)

$$r = r_M s_0/(s_0+k_m). \qquad (46)$$

But, we must note that, while (46) predicts a constant rate, (44) does not. Moreover, by virtue of (45), *r* is never the *initial rate*. The initial rate is always zero. Indeed, (44) or (46) refers to the *maximum rate* for a given run and $r_M$ stands for the overall maximum of all such maxima. In terms of our variables, (44) reduces to

$$R = d\delta/d\tau = \gamma = K_1\beta / (K_1\beta + K_2 + 1), \qquad (47)$$

where *R* stands for the rate in our terms. Form (47) [equivalently (44)] has another immediate problem. Putting the expansion (28) for $\beta$ at the right hand part of (47), one can check that there exists a non-zero first-order term in $\gamma$. Hence, this form cannot respect (23) or (45), which tells that there is a maximum of $\gamma$ in the $\gamma$ - $\tau$ plot. On the other hand, if one goes along with (47) and argues that $\gamma$ remains stationary only in an approximate sense [$\gamma_{c1} \approx 0$ in (21)], the first-order factor associated with ($\tau$ - $\tau_c$) should be very small. This leads to

$$\mu_c = K_1(K_2 + 1) / (K_1\beta_c + K_2 + 1)^2 \ll 1. \qquad (48)$$

Since the second part of (48) is naturally less than unity, one notes from here that, either a large $\beta_c$ (or, approximately, $\beta_0$) or a large $K_2/K_1$ would suffice. Indeed, the first requirement here corresponds closely to condition (31*a*) and the second one exactly to (31*d*). Otherwise, we may like to satisfy $K_1\beta_c \gg (K_2 + 1)$. Calling the left side of (48) by $\mu_c$ and defining, like (43), the quantity

$$\mu_0 = K_1(K_2 + 1) / (K_1\beta_0 + K_2 + 1)^2. \qquad (49)$$

one can test the performance of this criterion. However, we should admit that no strict theoretical basis of (47) exists.

**3.4. MM Kinetics in Absence of QSSA.** From (16), we get an exact equation for the rate of product formation. Form (23), along with the validity of the QSSA condition (39), would imply

$$|d\gamma/d\tau| \approx 0 \qquad (50)$$

over a considerable region beyond $\tau_c$. This choice yields

$$R = d\delta/d\tau \approx \gamma_c = K_1\beta_c / (K_1\beta_c + K_2 + 1) \qquad (51)$$

by virtue of (18). The rate is thus constant and (51) is the correct form of MM kinetics[17], admitting a linear growth of the product up to a certain range, and is in keeping with our discussion below (30). A different look on this point involves the mass balance equation (8b). At $\tau_c$, it reads as $\beta_0 = \beta_c + \gamma_c + \delta_c$. But, beyond $\tau_c$, balance is provided by first order effects from (28) and (30) up to the range of validity of QSSA. Explicitly, one can write the conservation as

$$\beta_0 = [\beta_c - \gamma_c(\tau - \tau_c)] + \gamma_c + [\delta_c + \gamma_c(\tau - \tau_c)], \tag{52}$$

rendering $\gamma$ constant at $\gamma_c$. Indeed, such a linear decay of $\beta$ and the concomitant linear rise in $\delta$ is the hallmark of the MM kinetics as well as QSSA.

Even when QSSA is not valid, we can still use (51) as

$$R_m = K_1\beta_c / (K_1\beta_c + K_2 + 1). \tag{53}$$

This means, the maximum rate is *always* given by (53). Translated to the parent form, it yields

$$r_m \equiv r_c = r_M s_c / (s_c + k_m), \, r_M = k_2 e_0. \tag{54}$$

Therefore, one needs to note the point $\tau_c$ at which rate attains the maximum value. The substrate concentration at this point is measured. Then, (54) offers a way to obtain the reaction constants via the familiar Lineweaver-Burk (LB) plot. The greatest advantage here is that, one *need not care* about QSSA. Hence, MM kinetics may be freed from the domain of validity of QSSA. We shall see the success of this endeavor later.

## 4. RESULTS AND DISCUSSION

**4.1. Concentration Profiles and Validity of QSSA.** The most direct way to check the validity of QSSA is to examine the $\gamma - \tau$ plot. However, for different sets, we have very different scales to detect changes in $\beta$, $\gamma$ or $\delta$. So, we have used an additional scaling. In place of a variable $x$, we employ

$$x_S = (x - x_{\min}) / (x_{\max} - x_{\min}) \tag{55}$$

**Figure 1. Scaled concentration profiles for the complex ($\gamma_S$): Set 9 (1, red); set 23 (2, blue); Set 56 (3, magenta); Set 55 (4, olive). Curve 1 depicts validity of the QSSA, but curves 3 and 4 do not. Curve 2 holds an intermediate position.**

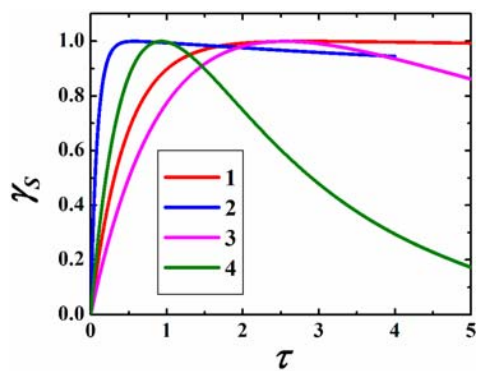

over the range under consideration. This does not affect the qualitative character of a plot, but different sets can be accommodated in the same graph. Figure 1 shows 4 representative plots.

Curve 1 is best in respect of the QSSA, curve 4 is worst. Case 3 also does not obey the QSSA and case 2 maintains an intermediate position. Based on such observations, one can classify various sets. The sets already mentioned in the figure correspond to the parameters summarized in Table 1.

Table 1. Behavior of a variety of sets defined by different reaction constants relative to $k_2$ ($k_2$ = 1, by choice) and with gradually lower substrate-enzyme ratios. The last column displays validity of QSSA for each set (Y: yes; N: no; I: intermediate), as evident from the concentration profiles of the complex.

| Set | $s_0$ | $e_0$ | $k_1$ | $k_{-1}$ | $t_c$ | $s_c$ | $c_c$ | Remark |
|---|---|---|---|---|---|---|---|---|
| 1 | 100 | 1 | 1/200 | 10 | 0.89 | 99.922 | 0.0434 | Y |
| 2 | 100 | 1 | 1/20 | 1 | 1.17 | 98.561 | 0.7113 | Y |
| 3 | 80 | 1 | 1/20 | 1 | 1.28 | 78.597 | 0.6627 | Y |
| 4 | 60 | 1 | 1/20 | 1 | 1.43 | 58.672 | 0.5946 | Y |
| 5 | 50 | 1 | 1/20 | 1 | 1.52 | 48.736 | 0.5492 | Y |
| 6 | 30 | 1 | 1/20 | 1 | 1.73 | 28.967 | 0.4200 | Y |
| 7 | 100 | 5 | 1/50 | 10 | 0.58 | 98.852 | 0.7617 | Y |
| 8 | 100 | 5 | 1/25 | 1 | 1.05 | 93.828 | 3.2618 | I |
| 9 | 100 | 5 | 1/500 | 1 | 2.86 | 98.468 | 0.4482 | Y |
| 10 | 20 | 1 | 1/10 | 1/10 | 1.82 | 18.427 | 0.6262 | I |
| 11 | 20 | 1 | 1/20 | 1 | 1.87 | 19.172 | 0.3240 | Y |
| 12 | 10 | 1 | 1/10 | 1/1000 | 2.20 | 8.713 | 0.4653 | N |
| 13 | 10 | 1 | 1/100 | 10 | 0.85 | 9.984 | 0.0090 | Y |
| 14 | 10 | 1 | 1/10 | 1 | 1.64 | 9.261 | 0.3165 | I |
| 15 | 10 | 1 | 1/100 | 1/1000 | 4.48 | 9.593 | 0.0874 | Y |
| 16 | 10 | 1 | 1/2 | 1/10 | 1.00 | 8.533 | 0.7950 | N |
| 17 | 10 | 1 | 1/5 | 1/10 | 1.59 | 8.599 | 0.6099 | I |
| 18 | 10 | 1 | 1/100 | 1/100 | 4.45 | 9.598 | 0.0868 | Y |
| 19 | 10 | 1 | 1/100 | 1/10 | 4.24 | 9.639 | 0.0806 | Y |
| 20 | 10 | 1 | 1/10 | 1/50 | 2.19 | 8.729 | 0.4611 | I |
| 21 | 10 | 1 | 1/100 | 1 | 2.92 | 9.837 | 0.0469 | Y |
| 22 | 10 | 1 | 1/100 | 20 | 0.51 | 9.993 | 0.0047 | Y |
| 23 | 10 | 1 | 1 | 1 | 0.58 | 8.783 | 0.8145 | I |
| 24 | 10 | 1 | 1 | 4/5 | 0.59 | 8.750 | 0.8294 | I |
| 25 | 10 | 1 | 1/10 | 1/500 | 2.20 | 8.713 | 0.4651 | N |
| 26 | 10 | 1 | 1/10 | 1/10 | 2.12 | 8.797 | 0.4444 | N |
| 27 | 10 | 1 | 1/20 | 1 | 2.02 | 9.490 | 0.1918 | I |

| Set | $s_0$ | $e_0$ | $k_1$ | $k_{-1}$ | $t_c$ | $s_c$ | $c_c$ | Remark |
|-----|-------|-------|-------|----------|-------|-------|-------|--------|
| 28 | 10 | 1 | 1/10 | 1/100 | 2.20 | 8.720 | 0.4633 | N |
| 29 | 8 | 1 | 1 | 4/5 | 0.65 | 6.773 | 0.7901 | N |
| 30 | 6 | 1 | 1 | 4/5 | 0.72 | 4.833 | 0.7286 | N |
| 31 | 5 | 1 | 1/10 | 1/10 | 2.30 | 4.224 | 0.2774 | I |
| 32 | 5 | 1 | 1/20 | 1 | 2.11 | 4.714 | 0.1054 | I |
| 33 | 5 | 1 | 1 | 4/5 | 0.76 | 3.888 | 0.6835 | N |
| 34 | 3 | 1 | 1 | 4/5 | 0.84 | 2.101 | 0.5386 | N |
| 35 | 10 | 5 | 1/10 | 1 | 1.09 | 7.480 | 1.3615 | N |
| 36 | 10 | 6 | 1/5 | 1 | 0.79 | 6.270 | 2.3124 | N |
| 37 | 10 | 8 | 1/30 | 2/3 | 1.49 | 7.712 | 1.0696 | N |
| 38 | 1 | 1 | 1 | 4/5 | 0.89 | 0.588 | 0.2462 | N |
| 39 | 1 | 1 | 1/100 | 10 | 0.85 | 0.998 | 0.0009 | Y |
| 40 | 1 | 1 | 1/100 | 20 | 0.51 | 0.999 | 0.0005 | Y |
| 41 | 1 | 1 | 1/100 | 1/1000 | 4.63 | 0.955 | 0.0094 | Y |
| 42 | 1 | 1 | 1/100 | 1/10 | 4.38 | 0.961 | 0.0086 | Y |
| 43 | 1 | 1 | 1/10 | 1/100 | 2.52 | 0.789 | 0.0725 | N |
| 44 | 1 | 1 | 1/10 | 1/10 | 2.43 | 0.806 | 0.0682 | N |
| 45 | 1 | 1 | 1 | 1/2 | 0.94 | 0.545 | 0.2665 | N |
| 46 | 1/2 | 1 | 1/10 | 1/10 | 2.45 | 0.400 | 0.0351 | N |
| 47 | 1/2 | 1 | 1 | 4/5 | 0.89 | 0.277 | 0.1336 | N |
| 48 | 5 | 10 | 1/50 | 1 | 1.53 | 4.134 | 0.3970 | N |
| 49 | 2/5 | 1 | 1 | 4/5 | 0.89 | 0.219 | 0.1086 | N |
| 50 | 1 | 10 | 1/1000 | 1/100 | 4.63 | 0.955 | 0.0094 | Y |
| 51 | 1 | 10 | 1/1000 | 1/10 | 4.40 | 0.960 | 0.0087 | Y |
| 52 | 1 | 10 | 1/1000 | 10 | 0.85 | 0.998 | 0.0009 | Y |
| 53 | 1 | 10 | 1/10 | 4/5 | 0.89 | 0.528 | 0.2851 | N |
| 54 | 1 | 10 | 1/100 | 10 | 0.64 | 0.990 | 0.0090 | Y |
| 55 | 1 | 10 | 1/10 | 1/2 | 0.93 | 0.480 | 0.3099 | N |
| 56 | 1 | 10 | 1/100 | 1/100 | 2.55 | 0.778 | 0.0764 | N |
| 57 | 5 | 100 | 1/500 | 1 | 1.54 | 4.106 | 0.4090 | N |
| 58 | 5 | 100 | 1/1000 | 10 | 0.64 | 4.930 | 0.0448 | Y |
| 59 | 5 | 100 | 1/10000 | 1/10 | 4.39 | 4.800 | 0.0436 | Y |
| 60 | 1 | 100 | 1/10000 | 1 | 2.99 | 0.9827 | 0.0049 | Y |

Indeed, Table 1 displays in a nutshell all the results of our numerical experiments. The relevant constants in terms of primitive symbols are given. The last column summarizes our observations on the validity of the QSSA [Y: yes; N: no; I: intermediate], based on features of the $\gamma - \tau$ plots. In this respect, we follow the outcomes of Figure 1 and classify the sets. Note that the sets vary widely in terms of the starting concentration ratios of the substrate and the enzyme, and the rate constants. We have maintained $k_2 = 1$ throughout and thus varied really the relative rate constants. This is what actually matters.

The table additionally shows the following general features: (i) It is *not* true that a low $t_c$ and a low $c_c$ are *necessary* for the satisfaction of QSSA, though such a condition may be *sufficient*. A comparison of sets 13 and 15 is worthwhile in this respect. (ii) It is also *not* true that a low value of $t_c$ would imply a low $c_c$, or *vice-versa*, even when QSSA is valid. Sets 39 – 42 would make the point clear. (iii) A high value of $t_c$ with moderately large $c_c$ may not *invalidate* the QSSA. Sets 15, 18 and 19 acknowledge this fact. (iv) If $s_0/e_0$ is large, it is *more* usual to find that QSSA is valid, unless $k_{-1} \ll k_1$. Only in the latter situation, one notices the breakdown. A number of sets would make the point clear. (v) When $e_0/s_0$ is large, invalidation of QSSA is commonplace. It is satisfied only if $k_{-1} \gg k_1$ is obeyed. (vi) The most complex case concerns the condition $s_0 \approx e_0$. Here, QSSA holds either with very small $k_1$ and the condition $k_{-1} < k_1$ or with moderate $k_1$ and the condition $k_{-1} > k_1$. Sets 38 – 45 provide evidence to such rationalizations.

We explore next whether the validity of QSSA has anything to do with the features of the $\beta - \tau$ plots. Figure 2 shows 4 such plots. One may note that curves 1 and 2 do show a faster linear fall-off initially, but the reduction becomes slower soon, though the decrease remains still linear. This is a hallmark if QSSA is valid. Otherwise, a less-than-linear decay is observed, as in case 4.

**Figure 2.** Scaled concentration profiles for the substrate ($\beta_S$): Set 58 (1, red); set 7 (2, blue); Set 59 (3, magenta); Set 53 (4, olive). Curves 1, 2 and 3 ensure validity of the QSSA, but curve 4 does not.

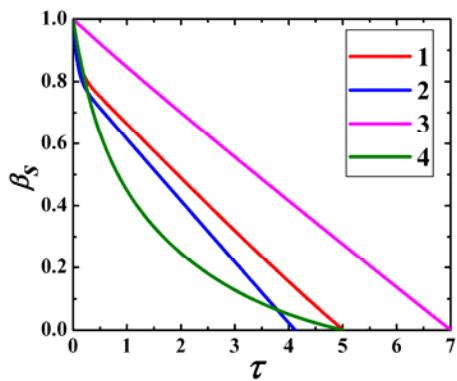

Curve 3 shows, on the contrary, a sharp, linear drop. QSSA is valid in this situation too. Similar plots are found for sets 15, 18, 19, 41, 42, 50 and 51. From Table 1, we detect that all such sets have high $t_c$, very small $K_1$ and small $K_2$. Under such conditions, we see from (26) that

$$d\beta/d\tau = -K_1\beta_0, \tau \to 0, \qquad (56)$$

that is valid even for reasonably large times, while (28) gives

$$d\beta/d\tau = -\gamma_c \qquad (57)$$

for $\tau$ around $\tau_c$ and well beyond, as required by the validity of QSSA. However, (18) shows that

$$\gamma_c \approx K_1\beta_c \approx K_1\beta_0. \qquad (58)$$

Hence, the two slopes coincide to give a single linear decay curve.

The $\delta$ - $\tau$ plot similarly contains signature of the validity of QSSA. Figure 3 shows again 4 plots. In accordance with our observations around (29), (30) and (52), we notice that $\delta$ starts with a quadratic rise but soon follows linearity. The linear region is large when QSSA is obeyed. In case QSSA fails to work, the growth rate gets reduced soon to yield a sigmoid profile. Cases 1 and 2 reveal typical linear regimes in support of QSSA; others show how such plots look when QSSA ceases to be obeyed.

**Figure 3. Scaled concentration profiles for the product ($\delta_S$): Set 39 (1, red); set 52 (2, blue); Set 38 (3, magenta); Set 45 (4, olive). Curves 1 and 2 ensure validity of the QSSA, but curves 3 and 4 do not.**

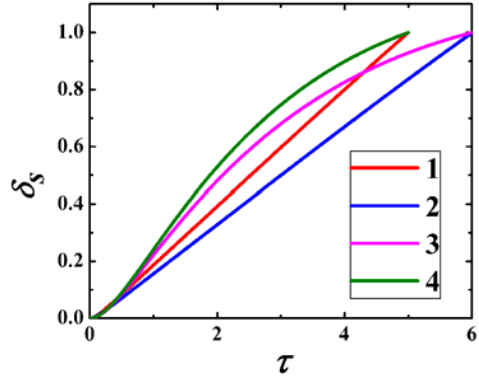

An alternative to the individual concentration plots is to go for triangular plots. We note from (8b) that

$$(\beta/\beta_0) + (\gamma/\beta_0) + (\delta/\beta_0) = 1. \qquad (59)$$

Therefore, calling these variables as $\beta_S$, $\gamma_S$ and $\delta_S$, we show the characteristics of situations that approve QSSA. Cases 1 and 4 in Figure 4 show complete breakdown of QSSA, whereas case 2 supports QSSA better than case 3. Indeed, after an initial rise from the right, if the line remains parallel to the $\beta_S$ axis, we note that QSSA is obeyed in such a case. In the best of cases, however, the plot looks much like a point.

**Figure 4. Reduced concentration profiles for substrate ($\beta_R$), complex ($\gamma_R$) and product ($\delta_R$) in triangular plots: Set 36 (1, red); set 54 (2, blue); Set 14 (3, magenta); Set 37 (4, olive). Curve 2 obeys best the QSSA and next comes curve 3, but curves 1 and 4 do not.**

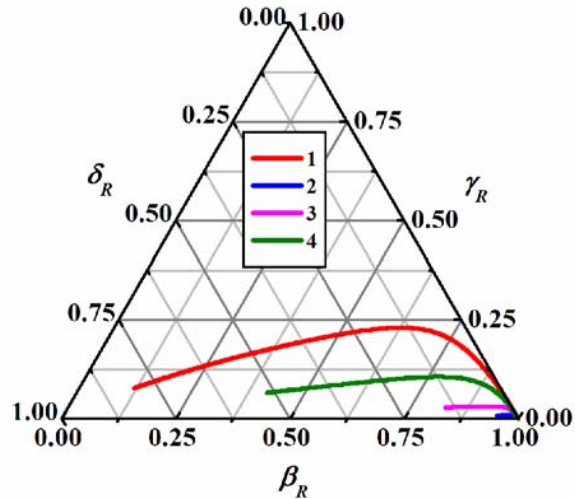

**4.2. Role of Timescales.** The validity of QSSA is often checked through timescales $\tau_c$ and $\tau_s$. The first one has a definite experimental basis, but the other one is a purely theoretical construct. So, we put such relations to test. Table 2 shows how the theoretical measure fares for the substrate-excess [s-QSSA] cases. The observed values are ordered and taken from Table 1,

**Table 2. Calculated and observed values of $\tau_c$ for a few sets with substrate in excess.**

| Set | $\tau_c$ (obs) | $\tau_c$ (calc*) |
|---|---|---|
| 34 | 0.84 | 0.21 |
| 13 | 0.85 | 0.09 |
| 4 | 1.43 | 0.20 |
| 37 | 1.49 | 0.50 |
| 11 | 1.87 | 0.33 |
| 43 | 2.52 | 0.90 |
| 21 | 2.92 | 0.48 |
| 15 | 4.48 | 0.91 |

*See eq. (36)

but the order breaks down miserably in case of calculated values. A similar problem is encountered with r-QSSA measure, as displayed in Table 3. As a result, some prediction based on (35) or (37) becomes misleading. This is precisely seen in Table 4. We find, whereas the adequacy of s-QSSA is rightly guessed by condition (35) with the expressions (36), the same of

Table 3. Calculated and observed values of $\tau_c$ for a few sets with enzyme in excess.

| Set | $\tau_c$ (obs) | $\tau_c$ (calc*) |
|---|---|---|
| 58 | 0.64 | 1 |
| 52 | 0.85 | 1 |
| 55 | 0.93 | 1 |
| 48 | 1.53 | 1 |
| 56 | 2.55 | 1 |
| 60 | 2.99 | 1 |
| 59 | 4.39 | 1 |
| 50 | 4.63 | 1 |

*See eq. (38)

Table 4. Prediction of the validity of s-QSSA and r-QSSA on the basis of the ratio $\tau_s/\tau_c$. The left column conforms to s-QSSA [see (35)], the right to r-QSSA [see (37)].

| Set | $\tau_s/\tau_c$ | Set | $\tau_s/\tau_c$ |
|---|---|---|---|
| 1 | 25287.36 | 56 | 10 |
| 2 | 979.02 | 57 | 5 |
| 3 | 718.56 | 58 | 10 |
| 4 | 500.00 | 59 | 100 |
| 5 | 405.41 | 60 | 100 |

r-QSSA does not correctly come out of (37), aided by (38). To check it explicitly, one may compare sets 56, 58, 59 and 60.

**4.3. Criteria to Justify QSSA.** We mentioned earlier that the various criteria for the validity of QSSA [*e.g.*, (31), (32) and (34)] do not stand[35] a close scrutiny. We have just seen that the conditions based on timescales also do not work unambiguously. So, it is now time to test the criteria that emerged from the present work. Specifically, we inquire about the smallness of $\lambda_c$ in (42) and $\mu_c$ in (48). However, these quantities involve $\beta_c$ that is not known *a priori*. So, we test, in addition, the performance-levels of $\lambda_0$ and $\mu_0$. Table 5 presents the results. A glance at it reveals that the QSSA is obeyed if both $\lambda_c$ and $\mu_c$ are less than 0.01. Roughly, the same criteria hold for $\lambda_0$ and $\mu_0$. On the other hand, when $\lambda_0, \mu_0 > 0.1$, one is sure that QSSA will not be obeyed at all. Cases for which $0.1 > \lambda_0, \mu_0 > 0.01$ show mostly a borderline behavior. This is how one can rationalize all our observations.

Table 5. Estimates of the characteristic constants $\lambda_0$ and $\lambda_c$ [see eqs. (43) and (42)], and $\mu_0$ and $\mu_c$ [see eqs. (49) and (48)], for all the sets quoted in Table 1. Low values predict the validity of QSSA.

| Set | $\lambda_0$ | $\lambda_c$ | $\mu_0$ | $\mu_c$ | Set | $\lambda_0$ | $\lambda_c$ | $\mu_0$ | $\mu_c$ |
|---|---|---|---|---|---|---|---|---|---|
| 1 | 0.0024 | 0.0024 | 0.0004 | 0.0004 | 31 | 0.0344 | 0.0361 | 0.0430 | 0.0475 |
| 2 | 0.0007 | 0.0007 | 0.0020 | 0.0021 | 32 | 0.0222 | 0.0224 | 0.0198 | 0.0200 |
| 3 | 0.0083 | 0.0084 | 0.0028 | 0.0028 | 33 | 0.1324 | 0.1582 | 0.0389 | 0.0556 |
| 4 | 0.0100 | 0.0101 | 0.0040 | 0.0041 | 34 | 0.1875 | 0.2307 | 0.0781 | 0.1183 |
| 5 | 0.0111 | 0.0113 | 0.0049 | 0.0051 | 35 | 0.1667 | 0.1820 | 0.1111 | 0.1324 |
| 6 | 0.0143 | 0.0145 | 0.0082 | 0.0084 | 36 | 0.3000 | 0.3688 | 0.1500 | 0.2267 |
| 7 | 0.0423 | 0.0424 | 0.0065 | 0.0065 | 37 | 0.1111 | 0.1155 | 0.1111 | 0.1201 |
| 8 | 0.0333 | 0.0348 | 0.0111 | 0.0120 | 38 | 0.3214 | 0.3769 | 0.2296 | 0.3156 |
| 9 | 0.0045 | 0.0046 | 0.0041 | 0.0041 | 39 | 0.0050 | 0.0050 | 0.0009 | 0.0009 |
| 10 | 0.0177 | 0.0187 | 0.0114 | 0.0127 | 40 | 0.0050 | 0.0050 | 0.0005 | 0.0005 |
| 11 | 0.0167 | 0.0169 | 0.0111 | 0.0114 | 41 | 0.0050 | 0.0050 | 0.0098 | 0.0098 |
| 12 | 0.0250 | 0.0267 | 0.0250 | 0.0286 | 42 | 0.0050 | 0.0050 | 0.0089 | 0.0089 |
| 13 | 0.0050 | 0.0050 | 0.0009 | 0.0009 | 43 | 0.0455 | 0.0464 | 0.0820 | 0.0852 |
| 14 | 0.0333 | 0.0342 | 0.0222 | 0.0234 | 44 | 0.0458 | 0.0466 | 0.0764 | 0.0789 |
| 15 | 0.0045 | 0.0046 | 0.0083 | 0.0083 | 45 | 0.3000 | 0.3667 | 0.2400 | 0.3587 |
| 16 | 0.0451 | 0.0512 | 0.0148 | 0.0191 | 46 | 0.0478 | 0.0482 | 0.0832 | 0.0846 |
| 17 | 0.0355 | 0.0390 | 0.0229 | 0.0277 | 47 | 0.3913 | 0.4333 | 0.3403 | 0.4173 |
| 18 | 0.0045 | 0.0046 | 0.0082 | 0.0083 | 48 | 0.0952 | 0.0960 | 0.0907 | 0.0922 |
| 19 | 0.0046 | 0.0046 | 0.0076 | 0.0077 | 49 | 0.4091 | 0.4458 | 0.3719 | 0.4416 |
| 20 | 0.0252 | 0.0272 | 0.0250 | 0.0285 | 50 | 0.0050 | 0.0050 | 0.0099 | 0.0099 |
| 21 | 0.0048 | 0.0048 | 0.0045 | 0.0045 | 51 | 0.0050 | 0.0050 | 0.0091 | 0.0091 |
| 22 | 0.0050 | 0.0050 | 0.0005 | 0.0005 | 52 | 0.0050 | 0.0050 | 0.0009 | 0.0009 |
| 23 | 0.0833 | 0.0927 | 0.0139 | 0.0172 | 53 | 0.4737 | 0.4858 | 0.4986 | 0.5243 |
| 24 | 0.0763 | 0.0853 | 0.0129 | 0.0162 | 54 | 0.0500 | 0.0500 | 0.0091 | 0.0091 |
| 25 | 0.0250 | 0.0267 | 0.0250 | 0.0286 | 55 | 0.4688 | 0.4845 | 0.5859 | 0.6260 |
| 26 | 0.0262 | 0.0278 | 0.0249 | 0.0281 | 56 | 0.0495 | 0.0496 | 0.0971 | 0.0975 |
| 27 | 0.0200 | 0.0202 | 0.0160 | 0.0163 | 57 | 0.0995 | 0.0996 | 0.0990 | 0.0992 |
| 28 | 0.0251 | 0.0268 | 0.0250 | 0.0285 | 58 | 0.0500 | 0.0500 | 0.0091 | 0.0091 |
| 29 | 0.0918 | 0.1050 | 0.0187 | 0.0245 | 59 | 0.0050 | 0.0050 | 0.0091 | 0.0091 |
| 30 | 0.1154 | 0.1357 | 0.0296 | 0.0409 | 60 | 0.0050 | 0.0050 | 0.0050 | 0.0050 |

**4.4. MM Kinetics and LB Plots.** We shall now see the effectiveness of the LB plots of the MM equations (53) or (54). The first set of data refers to sets 2 – 6, 11, 27 and 32. Here, only the last few sets show intermediate behavior with respect to QSSA. From the parameters of the concerned sets given in Table 1, one notes that the LB equation takes the form

$$1/R_m = 1/r_m = 1 + (k_m/s_c), \quad r_M = 1. \tag{60}$$

**Figure 5. LB plots for the sets 2, 3, 4, 5, 6, 11, 27 and 32 corresponding to choices (a) initial substrate concentration and (b) substrate concentration at $t_c$. All such sets have the same reaction constants and, except for the last few, they obey QSSA.**

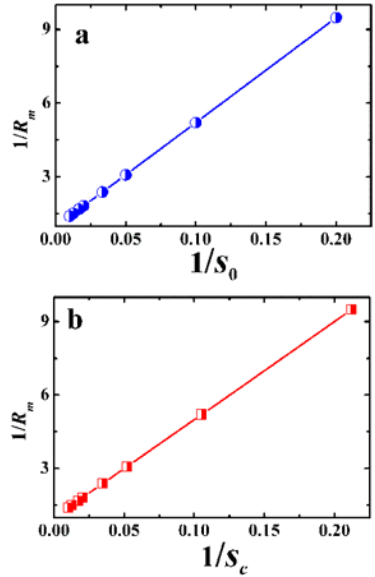

Conventional plots, however, employ $s_0$ in place of $s_c$, as shown in Figure 5(a). The data-set makes it also clear that the slope should turn out to be $k_m = 40$. Figure 5 shows the plots with two choices, the latter being ours. A least-squares-fit yields the following results:

Case (a): slope = 42.519; intercept = 0.971;

Case (b): slope = 40.043; intercept = 0.997.

We note happily that our case (b) offers much better values than the conventional plot (a).

The second set of data relates the sets 24, 29, 30, 33, 34, 38, 47 and 49. None of the sets chosen here obey the QSSA. The first few satisfy QSSA intermediately, but most of them violate the same badly. To notice this readily, one may consult Table 5 for values of both $\lambda_c$ and $\mu_c$ of the sets under study in Figures 5 and 6. Normally, for sets under consideration in Figure 6, one never goes for LB plots. But, if we are ready to fit an equation like (60), with $k_m = 9/5$, Figure 6 comes into sight. For the same two choices as above, one obtains here the results given below:

Case (a): slope = 3.344; intercept = 0.804;

Case (b): slope = 1.800; intercept = 1.000.

**Figure 6. LB plots for the sets 24, 29, 30, 33, 34, 38, 47 and 49 with choices (a) initial substrate concentration and (b) substrate concentration at $t_c$. All such sets have the same reaction constants and none obeys QSSA.**

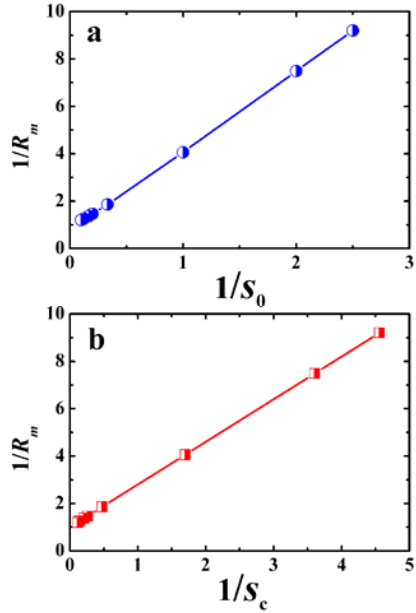

We thus see, our idea of employing the MM kinetics results in conditions defying QSSA is supported beyond any doubt. The agreements are indeed spectacular.

## 5. CONCLUSION

To summarize, we have studied here 60 sample cases with widely different reaction constants and substrate-enzyme ratios to check the conditions of validity of the QSSA. Our scheme is simple, but efficient. We have found that a reduced concentration profile of either the substrate [Figure 2] or the product [Figure 3] can also identify whether a given enzyme-substrate system obeys QSSA. This should be particularly useful to experimentalists because more often the concentration profile of the complex is difficult to follow. We have additionally found the theoretical importance of triangular plots [Figure 4] in deciphering a case of QSSA. In view of the limited success[35] of a number of prevalent criteria to check *a priori* the adequacy of QSSA, two new measures [Eqs. (42), (43) and (48), (49)] have been put forward. They emerged from our analysis. We checked thoroughly their efficacy [Table 5] and found them quite satisfactory. Most importantly, we have established that LB plots corresponding to the MM kinetics equations can be wisely employed to find the reaction constants even when QSSA ceases to hold. Figure 6 and the corresponding results establish our assertion beyond doubt.

It may be mentioned that several numerical approaches to study the reaction scheme (1) exist and a recent exposition[43] highlights quite a few earlier works. Certain endeavors[44-45] consider a catalytic cycle to handle (1). A detailed analysis[44] by casting the relevant equations in terms of a single nonlinear second order differential equation reveals some interesting features of the problem. Particularly notable is a subsequent work[45] that led to the emergence of an equation for the substrate concentration profile without invoking QSSA. However, the goals of such formulations differ from ours. MM kinetics is also of interest in stochastic simulation studies[46-48], electrocatalysis[49], etc. The relevance of QSSA[50] in such a context has drawn attention as well. We hope that the present endeavor may be useful in these backgrounds.

**Acknowledgement:**

**References:**

1. Flach, E. H.; Schnell, S. *IEE Proc. Syst. Biol.* **2006**, *153,* 187.
2. Pedersen, M. G.; Bersani, A. M.; Bersani, E.;Cortese, G. *MATCOM* **2008**, *79,* 1010.
3. Pedersen,M. G.; Bersani, A. M.; Bersani, E. *J. Math. Chem.* **2008**, *43,* 1318.
4. Tsodikova, S. G.; Shkel, I. A.; Tsodikov,O. V. *Analytical Biochemistry* **2009**, *387*, 276.
5. Pedersen, M. G.; Bersani, A. M. *J. Math. Biol.* **2010**, *60*, 267.
6. Michaelis, L.; Menten, M. L. *Biochem. Z.* **1913**, *49*, 333.
7. Briggs, G. E.; Haldane, J. B. S. *Biochem. J.* **1925**, *19*, 338.
8. Johnson, K. A.; Goody, R. S. *Biochem.* **2011**, *50*, 8264.
9. Hommes, F. A. *Arch. Biochem. Biophys.* **1962**, *96*, 28.
10. Bowen, J.; Acrivos, A; Oppenheim, A. *Chem. Eng. Sci.* **1963**, *18*, 177.
11. Wong, J. T. *J. Am. Chem. Soc.* **1965**, *87,* 1788.
12. Stayton, M. M.; Fromm, H. J. *J. Theor. Biol.* **1979**, *78*, 309.
13. Segel, L. A. *Bull. Math. Biol.* **1988**, *50,* 579.
14. Hanson, S. M.; Schnell, S. *J. Phys. Chem. A* **2008**, *112*, 8654.
15. Sanft, K. R.; Gillespie, D. T.; Petzold, L. R. *IET Syst. Biol.* **2011**, *5*, 58.
16. Bersani, A. M.; Dell'Acqua, G. *J. Math. Chem.* **2012**, *50*, 335.
17. Bajer, Z.; Strehler, E. *Biochem. Biophys. Res. Comm.* **2012**, *417*, 982.
18. Segel, L. A.; Slemrod, M. *SIAM Review* **1989**, *31*, 446.
19. Borghans, J. A. M.; De Boer, R. J.; Segel, L. A. *Bull. Math. Biol.* **1996**, *58*, 43.
20. Schnell, S.; Mendoza, C. *J. Theor. Biol.* **1997**, *189,* 207.
21. Schnell, S.; Maini, P. K. *Bull. Math. Biol.* **2000**, *2*, 483.
22. Tzafriri, A. R. B*ull. Math. Biol.* **2003**, *65*, 1111.
23. Tzafriri, A. R.; Elderman, E. R. *J. Theor. Biol.* **2004**, *226*, 303.
24. Tzafriri, A. R.; Elderman, E. R. *J. Theor. Biol.* **2007**, *245*, 737.
25. Kargi, F. *Biochem. Biophys. Res. Comm.* **2009**, *382*, 57.
26. Laidler, K. J. *Can. J. Chem.* **1955**, *33*, 1614.
27. Fraser, S. J. *J. Chem. Phys.* **1998**, *109*, 411.

28. Dell'Acqua, G.;  Bersani, A. M. *J. Math. Chem.* **2012**, *50,* 1136.
29. He, J. H. *Int. J. Nonlinear Mech.* **1999**, *34*, 699.
30. He, J. H.; Wu, X. H. *Chaos Solitons Fractals* **2006**, *29,* 108.
31. He, J. H. *J. Comput. Appl. Math.* **2007**, *207*, 3.
32. Meena, A.; Eswari, A.; Rajendran, L. *J. Math. Chem.* **2010**, *48*, 179.
33. Hanson, S. M.; Schnell, S. *J. Phys. Chem. A* **2008**, *112*, 8654.
34. Hill, C. M.; Waight, R. D.; Bardsley, W. G. *Mol. Cell. Biochem.* **1977**, *15*, 173.
35. Dhatt, S.; Bhattacharyya, K. *J. Math. Chem.* **2013** DOI 10.1007/s10910-013-0160-9.
36. Bispo, J. A. C.;  Bonafe, C. F. S.;  Koblitz, M. G. B.;  Silva, C. G. S.;  de Souza, A. R. *J. Math. Chem.* **2013**, *51,* 144.
37. Murray, J. D. *Mathematical Biology;* Springer, New York, **2002**.
38. Baker, G. A. Jr.; Graves-Morris, P. *Padé Approximants*, Cambridge University Press, Cambridge, **1996**.
39.  Bender, C. M.; Orszag, S. A. *Advanced Mathematical Methods for Scientists and Engineers,* Springer, New York **1999**.
40. Dhatt, S.; Bhattacharyya, K. *Int. J. Quantum Chem.* **2013**, *113,* 916.
41. Ouellet, L.; Laidler, K. J. *Can. J. Chem.* **1956**, *34,* 146.
42. Schnell, S.; Maini, P. K.; *Math. Comput. Modelling* **2002**, *35,* 137.
43. Gonzalez-Parra, G.; Acedo, L.; Arenas, A. *Comput. Appl. Math*. **2011**, *30*, 445.
44. Berberan-Santos, M. N. *MATCH Commun. Math. Comput. Chem*. **2010**, *63***,** 283.
45. Baleizão, C.; Berberan-Santos, M. N. *J. Math. Chem.* **2011**, *49*, 1949.
46. Mastny, E. A.; Haseltine, E. L.; Rawlings, J. B. *J. Chem. Phys*. **2007**, *127*, 094106.
47. MacNamara, S.; Bersani, A. M.; Burrage K; Sidje, R. B. *J. Chem. Phys*. **2008**, *129*, 095105.
48. Park, S.; Agmon, N.  *J. Phys. Chem.* B **2008**, *112*, 5977.
49. Puida, M.; Malinauskas, A.; Ivanauskas, F.  *J. Math. Chem.* **2012**, *50*, 2001.
50. Grima, R. *Phys. Rev. Lett.* **2009**, *102*, 218103.